\documentstyle[aps,twocolumn]{revtex}
\newcommand{\downeq}{\mathop{\rm \sim}\limits}
\begin{document}
\twocolumn[\hsize\textwidth\columnwidth\hsize\csname
@twocolumnfalse\endcsname

\draft
\title{Boundary polarization in the six-vertex model}
\author{\\N.~M.~Bogoliubov, A.~V.~Kitaev, and M.~B.~Zvonarev
}
\address{Steklov Institute of Mathematics at St. Petersburg,
 Fontanka 27, St.\ Petersburg 191011, Russia}
\date{\today}

\maketitle
\begin{abstract}
    Vertical-arrow fluctuations near the boundaries in the
    six-vertex model on the two-dimensional $ N \times N $ square
    lattice with the domain wall boundary conditions are considered.
    The one-point correlation function (``boundary polarization'') 
    is expressed via the partition function of the model on a sublattice.
    The partition function is represented in terms of standard objects
    in the theory of orthogonal polynomials.
    This representation is used to study the large $ N $ limit:
    the presence of the boundary affects the
    macroscopic quantities of the model even in this limit.
    The logarithmic terms obtained are compared with predictions
    from conformal field theory.

\end{abstract}
]

\section{The model}

 In this paper, we shall consider the six-vertex model
 on a  square lattice. Originally, this model was introduced as a model
 describing the ferroelectric properties of the hydrogen-bonded
 planar crystals \cite{LW}.
 The hydrogen atom positions are specified by attaching arrows to the
 lattice edges.
 In the six-vertex case, the arrows  are arranged in such a way
 that there are always two arrows pointing away from, and two arrows
 pointing into, each lattice vertex (the so-called ``ice rule'');
 thus, there are six possible states at each vertex (see, Fig.~\ref{vertices}).
 The statistical weights, $ {\sf a} $, $ {\sf b} $ and $ {\sf c} $,
 of the allowed states are invariant under the simultaneous reversal
 of all arrows:

\unitlength=1mm
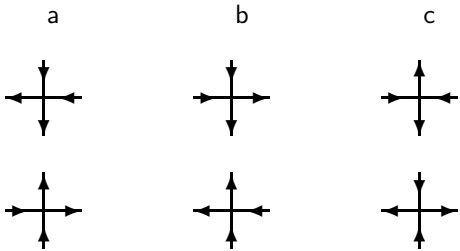
\begin{figure}[hbt]
\begin{center}
    \begin{picture}(60,35)
    \thicklines
%
% % % % % % % % % % % % % % %     vertex a1     % % % % % % % % % % % % % % %
%
        \put(5,15){\line(0,1){10}}
        \put(5.05,25){\vector(0,-1){3}}
        \put(5.05,20){\vector(0,-1){5}}
        \put(0,20){\line(1,0){10}}         %   ----->
        \put(5,20){\vector(-1,0){5}}
        \put(10,20){\vector(-1,0){3}}
        \put(5.5,30){$ {\sf a} $}
%
% % % % % % % % % % % % % % %     vertex a2     % % % % % % % % % % % % % % %
%
        \put(5,0){\line(0,1){10}}
        \put(5.05,5){\vector(0,1){5}}
        \put(5.05,0){\vector(0,1){3}}
        \put(0,5){\line(1,0){10}}              %   ----->
        \put(0,5){\vector(1,0){3}}
        \put(5,5){\vector(1,0){5}}
%
% % % % % % % % % % % % % % %     vertex b1     % % % % % % % % % % % % % % %
%
        \put(30,15){\line(0,1){10}}
        \put(30,25){\vector(0,-1){3}}
        \put(30,20){\vector(0,-1){5}}
        \put(25,20){\line(1,0){10}}         %   ----->
        \put(25,20){\vector(1,0){3}}
        \put(30,20){\vector(1,0){5}}
        \put(30.5,30){$ {\sf b } $}
%
% % % % % % % % % % % % % % %     vertex b2     % % % % % % % % % % % % % % %
%
        \put(30,0){\line(0,1){10}}
        \put(30,5){\vector(0,1){5}}
        \put(30,0){\vector(0,1){3}}
        \put(25,5){\line(1,0){10}}              %   ----->
        \put(30,5){\vector(-1,0){5}}
        \put(35,5){\vector(-1,0){3}}
%
% % % % % % % % % % % % % % %     vertex c1     % % % % % % % % % % % % % % %
%
        \put(55,15){\line(0,1){10}}
        \put(55,20){\vector(0,1){5}}
        \put(55,20){\vector(0,-1){5}}
        \put(50,20){\line(1,0){10}}         %   ----->
        \put(50,20){\vector(1,0){3}}
        \put(60,20){\vector(-1,0){3}}
        \put(55.5,30){$ {\sf c} $}
%
% % % % % % % % % % % % % % %     vertex c2     % % % % % % % % % % % % % % %
%
        \put(55,0){\line(0,1){10}}
        \put(55,10){\vector(0,-1){3}}
        \put(55,0){\vector(0,1){3}}
        \put(50,5){\line(1,0){10}}              %   ----->
        \put(55,5){\vector(-1,0){5}}
        \put(55,5){\vector(1,0){5}}
   \end{picture}
\end{center}
\caption{The vertices of the six-vertex model and their statistical weights.}
\label{vertices}
\end{figure}
\noindent
 The partition function of the model  on an $ N \times N $ square lattice is 
 obtained by summing over all possible arrow configurations $ \{C\}$,
    $$
        Z_N = \sum_{ \{C\} }^{} \,
        {\sf a}^{n_1} {\sf b}^{n_2} {\sf c}^{n_3},
    $$
 where $ n_1$, $ n_2 $, and $ n_3 $ are the number of vertices of type
 $ {\sf a} $, $ {\sf b} $, and $ {\sf c} $ in configuration $ C $,
 respectively ($ n_1 + n_2 + n_3 = N^2 $).

 The six-vertex model was studied for both periodic (PBC) \cite{LS,B}
 and fixed boundary conditions \cite{BBRY}.
 In this paper, we are concerned exclusively with the 
 domain wall boundary conditions (DWBC), namely, all arrows on the top
 and bottom of the lattice are pointing inward while all arrows
 on the left and right boundaries are pointing outward (see, Fig.~\ref{DWBC}).
 \unitlength=1.75pt
 \begin{figure}[h]
 \begin{center}
    \begin{picture}(60,60)
        \put(10,0){\line(0,1){60}}
        \put(10,0){\vector(0,1){7}}
        \put(10,60){\vector(0,-1){7}}
        \put(20,0){\line(0,1){60}}
        \put(20,0){\vector(0,1){7}}
        \put(20,60){\vector(0,-1){7}}
        \put(30,0){\line(0,1){60}}
        \put(30,0){\vector(0,1){7}}
        \put(30,60){\vector(0,-1){7}}
        \put(40,0){\line(0,1){60}}
        \put(40,0){\vector(0,1){7}}
        \put(40,60){\vector(0,-1){7}}
        \put(50,0){\line(0,1){60}}
        \put(50,0){\vector(0,1){7}}
        \put(50,60){\vector(0,-1){7}}
        \put(0,10){\line(1,0){60}}              %   ----->
        \put(10,10){\vector(-1,0){7}}
        \put(50,10){\vector(1,0){7}}
        \put(0,20){\line(1,0){60}}              %   ----->
        \put(10,20){\vector(-1,0){7}}
        \put(50,20){\vector(1,0){7}}
        \put(0,30){\line(1,0){60}}              %   ----->
        \put(10,30){\vector(-1,0){7}}
        \put(50,30){\vector(1,0){7}}
        \put(0,40){\line(1,0){60}}              %   ----->
        \put(10,40){\vector(-1,0){7}}
        \put(50,40){\vector(1,0){7}}
        \put(0,50){\line(1,0){60}}              %   ----->
        \put(10,50){\vector(-1,0){7}}
        \put(50,50){\vector(1,0){7}}
   \end{picture}
 \end{center}
 \caption{The domain wall boundary conditions.}
 \label{DWBC}
 \end{figure}
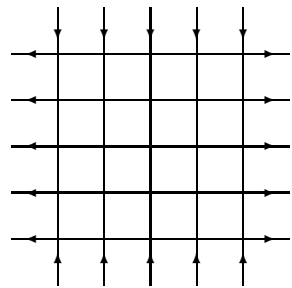

 \noindent 
 This model was introduced in Ref. \cite{K} in connection with
 the calculation of the correlation functions for exactly
 solvable 1+1 dimensional models \cite{KBI}. It appears that some
 problems from the theory of alternating sign matrices \cite{Kup,Br} and
 domino tilings \cite{JPS}
 may be reformulated in terms of this model.

 Aperiodic boundary conditions are of interest since they demonstrate
 the influence of the boundaries and internal defects of real physical
 systems on their bulk properties.
 By renormalization group method, it was shown that the behavior of the
 correlation functions near the surfaces and defects is quite different 
 from the bulk behavior \cite{HK}. The exactly solvable six-vertex model with 
 the DWBC provides us the opportunity to study 
 the surface phenomena beyond renormalization group scheme.

%%%%%%%%%%%%%%%%%%%%%%%%%%%%%%%%%%%%%%%%%%%%%%%%%%%%%%%%%%%%%%%%%%%%%%%%%%%%%
\section{The partition function and the boundary polarization}
 The partition function $ Z_{N} $ of the model with the DWBC 
 may be represented  as the determinant of an
    $ N \times N $ matrix \cite{I,ICK}.
 Though there exist several combinatorial representations
 for this determinant \cite{L,Kratt},
 it has so far only been calculated explicitly
 for some special cases \cite{Kup}.
 Significant progress has recently been achieved
 in studying the asymptotics of $ Z_N $ as $ N \to \infty $.
 Under some special restrictions on the values of the vertex weights,
 the bulk free energy was calculated in Ref. \cite{KZ-J}
 by using the Toda equation \cite{S}. A more general result was
 obtained in Ref. \cite{Z-J} by a reformulation of the six-vertex model
 as a Hermitian matrix model to which saddle point integration method
 was applied. A representation convenient for the large $N$ analysis
 was suggested also in Ref. \cite{NikS}.

 Less is known about the correlation functions of this model for at
 least two reasons. First, the calculation of the correlators,
 in general, is a more complicated problem than the calculation of
 the corresponding partition function. Second, the lack of
 translation invariance caused by the special boundary conditions
 introduces additional difficulties.
 Some correlation functions for the inhomogeneous
 model (the model with the statistical weights
 depending on the position of the vertex) 
 with special choice of the weights were considered in Ref. \cite{GK}.

 The boundary polarization is the one-point correlation function that
 describes the probability for the arrow on the fixed lattice edge
 on the boundary to be pointing in either direction.
 The symmetry of the model allows us to consider only vertical arrows.
 Let us denote by $ \chi_N $ the probability for the vertical arrow to be
 pointing down. Note that $ \chi_N $ for PBC  
 is independent of the position of the edge and is just a
 spontaneous polarization of the system \cite{B}.
 Here, we shall discuss $ \chi_N $ for the edge located at the lower-right
 corner of the lattice (in Fig.~\ref{corrf} this edge is dotted):

 \unitlength=1.75pt
 \begin{figure}[tbh]
 \begin{center}
    \begin{picture}(60,60)
        \put(10,0){\line(0,1){60}}
        \put(10,0){\vector(0,1){7}}
        \put(10,60){\vector(0,-1){7}}
        \put(20,0){\line(0,1){60}}
        \put(20,0){\vector(0,1){7}}
        \put(20,60){\vector(0,-1){7}}
        \put(30,0){\line(0,1){60}}
        \put(30,0){\vector(0,1){7}}
        \put(30,60){\vector(0,-1){7}}
        \put(40,0){\line(0,1){60}}
        \put(40,0){\vector(0,1){7}}
        \put(40,60){\vector(0,-1){7}}
        \put(50,0){\line(0,1){10}}
        \put(50,20){\line(0,1){40}}
        \put(50,0){\vector(0,1){7}}
        \put(50,60){\vector(0,-1){7}}
        \put(0,10){\line(1,0){60}}              %   ----->
        \put(10,10){\vector(-1,0){7}}
        \put(50,10){\vector(1,0){7}}
        \put(0,20){\line(1,0){60}}              %   ----->
        \put(10,20){\vector(-1,0){7}}
        \put(50,20){\vector(1,0){7}}
        \put(0,30){\line(1,0){60}}              %   ----->
        \put(10,30){\vector(-1,0){7}}
        \put(50,30){\vector(1,0){7}}
        \put(0,40){\line(1,0){60}}              %   ----->
        \put(10,40){\vector(-1,0){7}}
        \put(50,40){\vector(1,0){7}}
        \put(0,50){\line(1,0){60}}              %   ----->
        \put(10,50){\vector(-1,0){7}}
        \put(50,50){\vector(1,0){7}}
 \linethickness{0.5mm}
         \put(50,12){\line(0,1){2}}
         \put(50,16){\line(0,1){2}}
   \end{picture}
 \end{center}
 \caption{$ \chi_N $ is calculated for the arrow on the dotted edge. }
 \label{corrf}
 \end{figure}
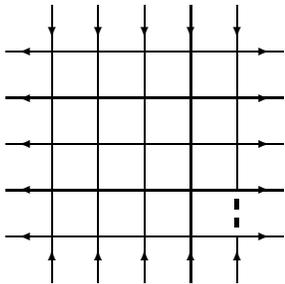
 \noindent
 It turns out that if the arrow on the dotted edge is pointing down,
 then due to the imposed boundary conditions, the allowed vertex configuration
 at the bottom and the right boundaries is determined in a unique way.
 Actually, in the lower-right corner one has the vertex of type $ {\sf c} $;
 thus, the rest of the $ 2N-2 $ vertices are of type $ {\sf b} $, and
 the DWBC are valid
 for the residual $ N-1 \times N-1 $ sublattice.
 Thus,
    \begin{equation}
        \chi_N = {\sf c \, b}^{2N-2} \frac{Z_{N-1}}{Z_N} ,
        \label{chi}
    \end{equation}
 and the problem
 of studying the correlation function, $ \chi_N $, is reduced to the analysis
 of the ratio of the partition functions $ Z_{N-1}/Z_{N} $.

 The boundary polarization on the arbitrary edge is expressed in terms of
 linear combination of partition functions on sublattices.
%%%%%%%%%%%%%%%%%%%%%%%%%%%%%%%%%%%%%%%%%%%%%%%%%%%%%%%%%%%%%%%%%%%%%%%%%%%%
\section{The connection with orthogonal polynomials}
 Due to the multiplicativity of the partition function, one can set
 $ \sf c=1 $ without loss of generality. Hence, the model is
 characterized by only two parameters, $ \sf a $ and $ \sf b $.
 In Fig.~4 the phase diagram on the
 $ ( {\sf a}, {\sf b} ) $ plane for the model with PBC is plotted
 (cf.,~Fig.~8.5 of Ref. \cite{B}).
 It may be regarded as the phase diagram for the model with the DWBC,
 in the sense that the free energy
 takes a different analytic form in the
 regions divided by the solid and dashed lines (see Ref. \cite{Z-J} for
 details).
 Naturally, the ground state and low temperature behavior do not coincide
 with those for the model with PBC.

 \unitlength=1pt
 \begin{figure}[tbh]
 \begin{center}
    \begin{picture}(120,140)
        \put(20,0){\line(0,1){120}}
        \put(20,120){\vector(0,1){14}}
        \put(120,20){\vector(1,0){14}}
        \put(0,20){\line(1,0){120}}              %   ----->
        \put(130,4){$ \sf a$}
        \put(0,130){$ \sf b$}
        \put(76,4){$ 1 $}
        \put(4,76){$ 1 $}
        \put(20,80){\line(1,1){10}}
        \put(80,20){\line(1,1){10}}
        \put(40,100){\line(1,1){10}}
        \put(100,40){\line(1,1){10}}
        \put(60,120){\line(1,1){10}}
        \put(120,60){\line(1,1){10}}
%%%%%%%%%%%%%%%%%%%%%%%%%%% Circle %%%%%%%%%%%%%%%%%%%%%%%%%%%%%%%%%%%%%%
        \put(80,20){\circle*{2}}
        \put(79.8,25.2){\circle*{2}}
        \put(79,30.4){\circle*{2}}
        \put(78,35.5){\circle*{2}}
        \put(76.4,40.5){\circle*{2}}
        \put(74.4,45.3){\circle*{2}}
        \put(72,50){\circle*{2}}
        \put(69.1,54.4){\circle*{2}}
        \put(66,58.5){\circle*{2}}
        \put(62.4,62.4){\circle*{2}}
        \put(58.5,66){\circle*{2}}
        \put(54.4,69.1){\circle*{2}}
        \put(50,72){\circle*{2}}
        \put(45.3,74.4){\circle*{2}}
        \put(40.5,76.4){\circle*{2}}
        \put(35.5,78){\circle*{2}}
        \put(30.4,79){\circle*{2}}
        \put(25.2,79.8){\circle*{2}}
        \put(20,80){\circle*{2}}
%%%%%%%%%%%%%%%%%%%%%%%%%%%%%%%%%%%%%%%%%%%%%%%%%%%%%%%%%%%%%%%%%%%%%%%%%
 \thicklines
        \put(80,20){\line(-1,1){60}}
        \put(79,20){\line(-1,1){59}}
   \end{picture}
 \end{center}
 \caption{ The phase diagram in terms of the weights $ {\sf a} $ and $
 {\sf b} $.}
 \label{diagram}
 \end{figure}
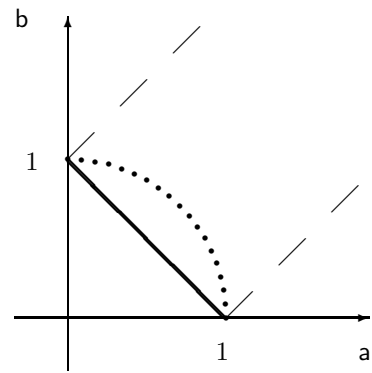
 In this and next section, we shall consider the model on the
 solid line in Fig.~\ref{diagram}.
 It is convenient to use the following parametrization
 of the vertex weights on this line:
     $$
         \left\{
         {{\sf a}= 1/2 - x\atop{\sf b}= 1/2 + x}
         \right. ,
         \qquad  -1/2 < x < 1/2 .
     $$

 It follows from Ref. \cite{I} that the partition function of the model
 is represented as the determinant of an $ N \times N $ matrix:
     \begin{equation}
         Z_{N} = \frac{\det_N {\cal M}}{[\phi(x)]^{N^2}
         \left( \prod\limits_{k=1}^{N-1} k! \right)^2} \, ,
     \label{statsum}
     \end{equation}
 where the matrix elements of $ {\cal M} $ are given by
     \begin{equation}
         {\cal M}_{\alpha k}=\frac{d^{\alpha + k}}{d x^{\alpha + k}} \phi(x) ,
         \qquad \alpha,k =0,1,\ldots,N-1 ,
         \label{M}
     \end{equation}
 and
     $
         \phi(x)=\left[(\frac1 2 - x)(\frac1 2 + x)\right]^{-1} .
     $

 Matrix $ {\cal M} $ is a Hankel matrix, that is,
 it has constant entries along the antidiagonals.
 Its determinant can be expressed in terms of objects related to
 the theory of orthogonal polynomials.
 Let $ p_{n} (\xi)$, $ n \ge 0 $, be a sequence of monic orthogonal
 polynomials with weight $ \rho (\xi) $:
     \begin{equation}
         \int_{-\infty}^{\infty}
         p_{n}(\xi) p_{m}(\xi) \rho(\xi) d \xi =
         \delta_{mn}h_{n},  \qquad  n,m \ge 0 .
     \label{h}
     \end{equation}
 Then the determinant of the Hankel matrix with the elements
     \begin{equation}
         {\cal H}_{\alpha k} = \int_{-\infty}^{\infty}
         \xi^{\alpha+k} \rho(\xi) d \xi ,
         \qquad \alpha, k = 0,1, \ldots, N-1 ,
     \label{H}
     \end{equation}
 is equal to \cite{Kratt}
     $
         \det\nolimits_N {\cal H} = h_{0}^N
         b_1^{N-1} b_2^{N-2} \cdots b_{N-2}^2 b_{N-1} 
     $.
 Here $ h_{0} $ is determined from Eq.~(\ref{h}), and
 $ b_n $ are the coefficients of the corresponding three-term
 recurrence relation,
     $$
         p_{n+1}(\xi) = (a_n + \xi) p_n (\xi) - b_n p_{n-1} (\xi) ,
         \qquad n \ge 1,
     $$
 with the initial conditions $ p_0 (\xi) = 1 $ and $ p_1 (\xi) = \xi + a_0 $.

 The function $ \phi(x) $ has the following integral representation:
     $$
         \phi(x) = \int_{-\infty}^{\infty}
         e^{-x \xi} e^{-|\xi|/2} d \xi .
     $$
 Combining this representation with Eq.~(\ref{M}) and Eq.~(\ref{H}),
 one shows that 
     $\det_N {\cal M}=\det_N {\cal H} $,
 where the weight $ \rho(\xi) $ is equal to
     \begin{equation}
         \rho(\xi) = e^{-x \xi} e^{-|\xi|/2} .
         \label{rho}
     \end{equation}
 Then, since the coefficients $ b_n $ and $ h_n $ 
 satisfy the well-known relation
     $ b_n=h_n/h_{n-1} $, $ n \ge 1 $,
 we have
     $ \det\nolimits_N {\cal M} = h_{0} h_{1} \cdots h_{N-1} $,
 and the desired representation  for probability (\ref{chi}) is \cite{BKZ}
    \begin{equation}
        \chi_{N+1}= {\sf a}^{-2N-1} {\sf b}^{-1} \, \frac{(N!)^2}{h_{N}} .
    \label{eq:difz}
    \end{equation}

 Now let us discuss briefly the general case when $ {\sf a} $ and $ {\sf b} $
 are arbitrary positive constants.
 The partition function is given by formulas (\ref{statsum}) and 
 (\ref{M}), while $ \phi(x) $
 should be changed (see, for example, \cite{KBI,I,ICK}),
 the weight $ \rho(\xi) $ is obtained by straightforward calculations
 \cite{Z-J}, and the corresponding expression for $ \chi_{N+1} $
 differs from Eq.~(\ref{eq:difz}) by a constant.
%%%%%%%%%%%%%%%%%%%%%%%%%%%%%%%%%%%%%%%%%%%%%%%%%%%%%%%%%%%%%%%%%%%%%%%%%%%%%%%
\section{The Large $ N $ limit}

 Equation (\ref{eq:difz}) reduces the problem of calculation of the
 probability $\chi_{N+1}$ to the calculation of the
 normalizing coefficient $h_{N}$.
 At present there exists a powerful method for studying the
 $ N \to \infty $ behavior of $ h_N $. This method is based on the
 matrix Riemann-Hilbert conjugation problem \cite{FIK}.
 The corresponding asymptotic technique was suggested in Ref. \cite{DKMVZ}
 and worked out for orthogonal polynomials with nonanalytic weights
 in Ref. \cite{KM}. Using this technique, we obtain the main result of
 the paper:
     \begin{eqnarray}
         \ln h_N = 2N \ln N +
         2N \ln \left[\frac{\pi e^{-1}}{\cos (\pi x)}\right] + \ln N
         \nonumber \\     
         + \ln \left[ \frac{2\pi^2}{\cos(\pi x)} \right]
         + \frac{1}{4N} 
         +\frac{\varphi(x,N)}{2N (\ln N)^2 } + \cdots ,
     \label{exp:h}
     \end{eqnarray}
 where
     $$
         \varphi(x,N) = (-1)^N \cos[2 \pi x (N+ 1/2)]
     $$
 and the omitted terms are of the order $ 1/[N (\ln N)^3] $. 
 Note that for $ x=0 $ this result coincides
 with the one obtained in Ref. \cite{KM}.

 Having expansion (\ref{exp:h}),
 it is easy to find the expansion for $ \chi_N $:
     \begin{eqnarray}
         &&\ln \chi_{N} =
         \nonumber\\
         &&-2N \ln \left[ \frac{\pi(1/2-x)}{\cos(\pi x)} \right]
         +\ln\left[\frac{\pi}{\cos(\pi x)} \frac{1/2-x}{1/2+x} \right]
         \nonumber\\
         && -\frac{1}{12N} 
         - \frac{\varphi(x,N-1)}{2N(\ln N)^2 } + \ldots .
     \label{expans.log}
     \end{eqnarray}
 From Eq.~(\ref{expans.log}), we get all increasing terms
 for the partition function $ Z_N $,
     \begin{equation}
         Z_N \downeq\limits_{N\to\infty}^{}
         C\exp\left( f_0 N^2 + f_1 N + f_2 \ln N \right) ,
     \label{Z_N}
     \end{equation}
 where $ C>0 $ is a bounded function of $ N $, and
     $$
         f_0= \ln\left[ \frac{\pi(1/4-x^2)}{\cos(\pi x)} \right],
         \quad f_1=0, \quad f_2=\frac{1}{12}.
     $$
%%%%%%%%%%%%%%%%%%%%%%%%%%%%%%%%%%%%%%%%%%%%%%%%%%%%%%%%%%%%%%%%%%%%%%%%%%%%%%
\section{Results at other points}

 At present there are several points on the phase diagram, where
 the determinant has been calculated in closed form.

 (i) The ``free-fermion case'', $ {\sf a}^2 + {\sf b}^2 =1 $
 (dotted circular quadrant on the phase diagram). On this circle, one has
 a very simple result $ Z_N =1 $. This result can be obtained
 by an appropriate limit from the inhomogeneous partition function
 \cite{KBI,ICK}. Therefore, one has
     \begin{equation}
         \ln \chi_N = 2(N-1) \ln {\sf b} \, .
     \label{chiferr}
     \end{equation}

 (ii) The ``ice point'', $ {\sf a} = {\sf b} =1 $. All weights are equal,
 and the partition function is just the number of
 allowed configurations $ \{C\} $  (Sec.\ I).
 In this case one has Ref. \cite{Kup}
     $$
         Z_N = \prod\limits_{j=1}^{N}
         \frac{(3j-2)!}{(N-1+j)!} ;
     $$
 thus,
     \begin{equation}
         \ln \chi_{N}  =
         N \ln\frac{16}{27} -\frac12 \ln \frac{16}{27} + \frac5{36} \frac1N
         +\frac{5}{72} \frac1{N^2} + \cdots .
     \label{chiice}
     \end{equation}
 For $ Z_N $ we have asymptotic expansion (\ref{Z_N}),
 where 
     $ f_0=\ln\frac{3\sqrt{3}}4 $, $ f_1=0 $ and $ f_2=-\frac5{36} $.
 Note that for the model with PBC \cite{B}, 
     $ f_{0}^{\rm PBC} = \ln \frac8{3\sqrt{3}} > f_{0} $, that is,
 for large $ N $ the number of allowed configurations for the DWBC is less than
 for periodic ones.

 (iii) The point  $ {\sf a} = {\sf b} =1/\sqrt{3} $. In this case \cite{Kup}
     $$
         Z_{2N+1} = \frac1{3^{N^2}} \left[ \prod\limits_{j=1}^{N}
         \frac{(3j-1)!}{(N+j)!} \right]^2 ,
     $$
     $$
         Z_{2N} = \frac{(3N-1)!(N-1)!}{3^N [(2N-1)!]^2} Z_{2N-1} ,
     $$
 and we immediately get
     \begin{equation}
         \ln\chi_{N} =
         N \ln\frac{4}{9} + \frac{1}{2}\ln\frac{27}{4} -\frac1{18} \frac1{N}
         + \cdots ;
     \label{chi3}
     \end{equation}
 thus, $ f_0= \ln\frac{\sqrt{3}}2 $, $ f_1=0 $ and $ f_2=\frac1{18} $. 
 The first omitted term in Eq.~(\ref{chi3}) is of the order $ N^{-2} $,
 and depends on the parity of $ N $.

 We complete this paper with the following statement.
 One can see that in expansion (\ref{expans.log}), which was obtained
 for $ {\sf a} + {\sf b} = 1 $, there exists a logarithmic term.
 For arbitrary $ {\sf a} $ and $ {\sf b} $,
 we state that the logarithmic terms in the large $ N $ expansion
 of $ \ln \chi_{N} $ will take place only under the following conditions:
 weights $ {\sf a} $ and $ {\sf b} $ are either on the solid line
    $ {\sf a} + {\sf b} = 1 $,
 or on the dashed lines
    $ | {\sf a} - {\sf b}| =1 $ of the phase diagram (Fig.~\ref{diagram}).
 For the model with PBC, the logarithmic terms
 could be explained via conformal field theory \cite{KBI}.
 Similar arguments are valid for the discussed model.
 The influence of the boundary condition changing operators \cite{C}
 should also now be taken into consideration.
 If we move away from the solid or dashed lines, the logarithmic terms 
 vanish and the expansion of $ \ln\chi_N $ will be in integer powers of $ N $.
 Expansions (\ref{chiferr}), (\ref{chiice}) and (\ref{chi3}) indeed
 contain only integer powers of $N$, thus confirming the above statement.

 Finally, we would like to mention that the six-vertex model with any
 boundary conditions can be considered as a model for a description
 of interface roughening of a crystal surface \cite{Be}. 
 An important point in these studies is the existence of exact
 analytical results, which are known for the six-vertex model with PBC
 \cite{LW,LS,B} .
 We believe that our analytical results for the model with DWBC provide
 one more basis for the experiments and simulations in this direction.
 
 We thank A.~H.~Vartanian for bringing Ref. \cite{KM} to our attention.
 This work was partially supported by the RFBR Grant No. 01-01-01045.

\end{document}